\documentclass[showkeys,preprint,twocolumn,11pt,times,superscriptaddress]{revtex4-2}
\usepackage[utf8]{inputenc}
\usepackage[english]{babel}
\usepackage{graphicx}
\usepackage{dcolumn} 
\usepackage{bm} 
\usepackage[colorlinks=true,
            linkcolor=blue,
            urlcolor=blue,
            citecolor=blue]{hyperref}
\usepackage[mathlines]{lineno} 
\usepackage{amsmath, amsfonts}
\usepackage{physics}
\usepackage{booktabs}
\usepackage{tabularx}
\usepackage{geometry}
\DeclareMathOperator{\sign}{sign}

\begin{document}

\title{Identification of single- and double-well coherence-incoherence patterns by the binary distance matrix}

\author{Vagner dos Santos}
\affiliation{Department of Mathematics and Statistics, State University of Ponta Grossa, 84030-900, Ponta Grossa, PR, Brazil}
\affiliation{Ci\^ encias Exatas, Naturais e Engenharias, Centro Universit\' ario UNIFATEB, 84266-010, Tel\^ emaco Borba, PR, Brazil}

\author{Matheus Rolim Sales}
\affiliation{Graduate Program in Sciences/Physics, State University of Ponta
Grossa, 84030-900, Ponta Grossa, PR, Brazil}
\affiliation{Potsdam Institute for Climate Impact Research, Member of the Leibniz Association, P.O. Box 6012 03, D-14412 Potsdam, Germany}
\affiliation{Institute of Mathematics, Humboldt University Berlin, 12489 Berlin, Germany}

\author{Sishu Shankar Muni}
\affiliation{Department of Physical Sciences, Indian Institute of Science Education and Research Kolkata, Mohanpur, West Bengal, 741246, India}

\author{Jos\'e D. Szezech Jr}
\affiliation{Department of Mathematics and Statistics, State University of Ponta Grossa, 84030-900, Ponta Grossa, PR, Brazil}
\affiliation{Graduate Program in Sciences/Physics, State University of Ponta
Grossa, 84030-900, Ponta Grossa, PR, Brazil}

\author{Antonio Marcos Batista}
\affiliation{Department of Mathematics and Statistics, State University of Ponta Grossa, 84030-900, Ponta Grossa, PR, Brazil}
\affiliation{Graduate Program in Sciences/Physics, State University of Ponta
Grossa, 84030-900, Ponta Grossa, PR, Brazil}

\author{Serhiy Yanchuk}
\affiliation{Potsdam Institute for Climate Impact Research, Member of the Leibniz Association, P.O. Box 6012 03, D-14412 Potsdam, Germany}
\affiliation{Institute of Mathematics, Humboldt University Berlin, 12489 Berlin, Germany}

\author{Jürgen Kurths}
\affiliation{Potsdam Institute for Climate Impact Research, Member of the Leibniz Association, P.O. Box 6012 03, D-14412 Potsdam, Germany}
\affiliation{Institute of Physics, Humboldt University Berlin, 10099 Berlin, Germany}
\affiliation{Division of Dynamics, Lodz University of Technology, Stefanowskiego 1/15, 90-924 Lodz, Poland}


\begin{abstract}
The study of chimera states or, more generally, coherence-incoherence patterns has led to the development of several tools for their identification and characterization. In this work, we extend the eigenvalue decomposition method to distinguish between single-well and double-well patterns. By applying our method, we are able to identify the following four types of dynamical patterns in a ring of nonlocally coupled Chua circuits and nonlocally coupled cubic maps:  single-well cluster, single-well coherence-incoherence pattern, double-well cluster, and double-well coherence-incoherence. In a ring-star network of Chua circuits, we investigate the influence of adding a central node on the spatio-temporal patterns. Our results show that increasing the coupling with the central node favors the occurrence of single-well coherence-incoherence states. We observe that the boundaries of the attraction basins resemble fractal and riddled structures.
\end{abstract}

\maketitle

\section{Introduction}
\label{sec:intro}

Since Huygen's discovery of synchronization \cite{Huy17}, coupled oscillators have became an important paradigm in nonlinear science. 
Numerous studies were devoted to synchronized and desynchronized oscillations. 
Kuramoto and Battogtokh \cite{kuramoto2002coexistence} observed a coexistence of synchronous and asynchronous oscillations \cite{PhysRevE.70.056125}, known as chimera state \cite{strogatz_chimera}, in a network of coupled phase oscillators. 
Since then chimera states have been studied in various systems ranging from chemical \cite{chimera_chemical1,chimera_chemical2,chimera_chemical3,Haugland2021}, electronic oscillators \cite{chimera_electronic1,chimera_electronic2,chimera_electronic3} to neuron systems \cite{SANTOS201786, chimera_neuron1,chimera_neuron2,chimera_neuron3}. Different classification schemes to identify chimera states were proposed \cite{omel2018mathematics,scholl2016synchronization}. Researchers have considered various topologies of networks to study the emergence of chimeras, such as ring-star \cite{Muni2020}, lattice \cite{ShMu20a,ShMu20b,OmelChenko2012}, multiplex networks of both discrete 
\cite{Muni22a,Muni22b} and continuous dynamical systems \cite{ShMu21a,ShMu21b,ShMu21c}. The effects of different coupling topologies have been investigated \cite{scholl2016synchronization,Martens_2016}. 
One motivation for such an exploration of topologies is that some neuronal dysfunctions might be associated with chimera states in the presence of certain network structures \cite{uhlhaas2006neural}. 
Chimera states were found in natural systems such as interacting fireflies or mechanical experiments \cite{Dudkowski2020}.

Meanwhile, many different patterns containing coherent and incoherent parts are often referred to as chimeras in the literature. Sometimes these patterns have no obvious relation to the chimera states originally described in \cite{kuramoto2002coexistence}, as in the case of our work. To avoid confusion, we use the more general term \textit{coherence-incoherence patterns} in this work.

Multistability is a phenomenon that appears in a large class of dynamical systems. It refers to the coexistence of multiple attractors (at least two) in the dynamical system for the same set of parameter values. Different initial conditions in the phase space then leads to different attractors. From a dynamical perspective, the presence of multiple attractors is a striking feature. There are many studies in the literature with applications of multistability in neuroscience \cite{Malashchenko2011}, optics \cite{Yanchuk2010a}, or engineering systems \cite{Feu18}. 
In an extreme case, infinitely many attractors can coexist \cite{Newhouse1974}. Recently, a detailed geometric mechanism behind such a phenomenon was studied in Refs.~\cite{Mu22,MuMcSi21}. 
The coexistence of two different coherence-incoherence states for the same set of parameter values was also observed. In a network of Chua's circuits \cite{shepelev2017chimera,Muni2020} and in a network of cubic maps \cite{shepelev2017chimera}, it was verified that both single-well (SW) and double-well (DW) coherence-incoherence states coexist for the same parameter set. In that case, different initial conditions can induce different states, such as SW and DW coherence-incoherence states \cite{Sh17a, Muni2020, shepelev2017chimera}.

The Chua's circuit is known for its simplicity in terms of its components. It contains one inductor, one diode, two capacitors, and one nonlinear resistor \cite{Chua92}. The Chua's circuit has been applied in secure communication systems \cite{Chua:secure}, Gaussian coloured noise \cite{Chua:noise}, and hand-written patterns recognition \cite{chua:recog}. It is also well known for its double-scroll attractor. Different spatiotemporal patterns, for instance, spiral waves, have been found in networks of coupled Chua's oscillators \cite{Chua:spiral}. As for the cubic map, it is the simplest map to present bistable dynamics and it can be thought as an analog of the Chua's circuit for discrete-time systems \cite{cubicmap1}. This map can demonstrate both regular and chaotic dynamics depending on its parameters, and the SW and DW coherence-incoherence states were observed in a network of coupled cubic maps with chaotic bistable dynamics \cite{shepelev2017chimera}.

The purpose of this work is threefold. i) The first is to identify and characterize two different coherence-incoherence states, as well as cluster states, using an extension of the method of eigenvalue decomposition \cite{Parastesh2020}. We use a network of Chua's circuits and a network of cubic maps to demonstrate our methodology. ii) The second purpose is to explore the coexistence of spatio-temporal patterns, and the iii) last one is to analyze the basins of attraction in the space of initial conditions. We study the coexistence of coherence-incoherence states in a network of Chua's circuits and compute the basins of attraction of each state present in the system for a set of parameters \cite{Va18}. 


The paper is organized as follows. Section \ref{sec:method} illustrates the method used to characterize different network states such as cluster synchronization, desynchronization, SW coherence-incoherence state, and DW coherence-incoherene state. Section \ref{sec:model} introduces the ring-star network of Chua circuits and the network of cubic maps. Section \ref{sec:result} discusses the existence of bistability of coherence-incoherence states in the Chua's network and section \ref{sec:concl} presents our final remarks.


\section{Methodology}
\label{sec:method}

Aiming to characterize the distinct dynamical states in a network of coupled oscillators, we extend the method of eigenvalue decomposition \cite{Parastesh2020}. We begin by numerically integrating the equations of motion, in case of a flow, or by iterating the difference equations, in case of a mapping, to obtain the state variables time series $\mathbf{x}_i(t)$. We then construct the symmetric spatial distance matrix $\mathbf{d}$, according to
\begin{equation}
    \label{eq:distmat}
    d_{ij} = \left\langle\|{x}_i(t) - {x}_j(t)\|\right\rangle_t,
\end{equation}
where $\|\cdot\|$ is the Euclidean norm, $\langle\cdot\rangle_t$ denotes the average in time, $i,j=1,2,\ldots,N$, and $N$ is the network size. If $d_{ij}\approx 0$, the nodes $i$ and $j$ are coherent and large values of $d_{ij}$ indicate incoherence. From $\mathbf{d}$, we define a binary matrix with elements equal to $1$ and $0$ denoting coherence and incoherence, respectively. These elements are computed as
\begin{equation}
    \label{eq:recmat}
    L_{ij}^{s} = \Theta\left(\delta_s - d_{ij}\right),
\end{equation}
where $\delta_s > 0$ is a small threshold and $\Theta$ is the Heaviside function. Defined this way, $L_{ij}^{s}$ equals $1$ if the absolute difference of $x_i$ and $x_j$ is smaller than $\delta_s$ and zero otherwise. This definition of $L_{ij}^s$ is similar to the definition of the recurrence matrix used in the recurrence quantification analysis \cite{Eckmann_1987, rec1, rec2, rec3, rec4, rec5}, however, the recurrence matrix is evaluated using the time series of the state variables and our matrix is evaluated using the state variables of the nodes in the network. Such definition is often called spatial recurrence plot \cite{Vasconcelos2006, SANTOS20152188}.

In the case of a completely incoherent state, the matrix $\mathbf{L}^{s}$ is the identity matrix and all its eigenvalues are equal to $1$. When some of the nodes are coherent, some off-diagonal elements of $\mathbf{L}^{s}$ are nonzero and there are eigenvalues greater than $1$ and eigenvalues less than $1$ in order to satisfy $\sum_k \lambda_k = \tr(\mathbf{L}^{s}) = N$ \footnote{The binary distance matrix has the properties of the correlation matrix},  where $\tr(\mathbf{L}^{s})$ is the trace of $\mathbf{L}^{s}$. Parastesh \textit{et al.} \cite{Parastesh2020} reported that these eigenvalues are related to the coherent nodes and the corresponding eigenvectors show the positions of the synchronized nodes in the network. By comparing the elements of the eigenvectors, one can obtain the number of incoherent nodes of the network.

The drawback of this methodology is that one cannot apply it to large networks because the computational cost of the evaluation of eigenvalues and eigenvectors of large matrices is too high. One can, however, use the properties of the eigenvalues and eigenvectors reported by Parastesh \textit{et al.} to simplify the method while getting the same results. Here we focus on the sum of the elements of each column of the binary matrix, $s_j = \sum_i L_{ij}^{s}$. If the $j^{\rm th}$ node is incoherent with the rest of the network, then $s_j = 1$, as all elements of this column are zero except for the diagonal element. If there is coherence with another node, then $s_j > 1$. In this way, one can detect the positions of the synchronized nodes in the network.
   

\section{Network models and pattern detection}
\label{sec:model}

In this study, we investigate the dynamics of a network of $N+1$ identical Chua's oscillators coupled in the ring-star topology and the dynamics of $N$ identical cubic maps coupled in the ring topology. First we focus on finding how to distinguish between the different dynamical behaviors present in the systems using our methodology. Then, we study how spatio-temporal patterns change as a coupling parameter is varied. 

\subsection{Chua circuits}

\begin{figure}[tb!]
    \centering
    \includegraphics[width=0.7\linewidth]{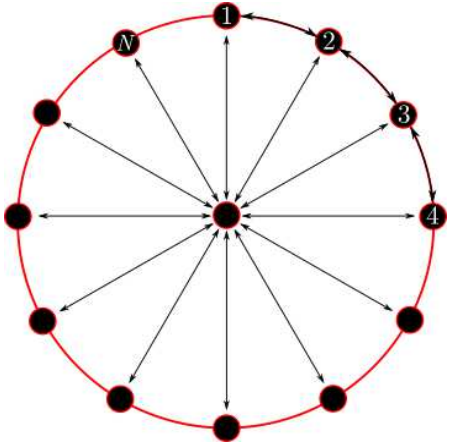}
    \caption{The ring-star network with $N$ nodes forming a circular ring plus one node at the center coupled according to a star topology. The ring nodes are labeled $i=1,2,\ldots,N$ and the central node is labeled $i=N+1$.}
    \label{fig:ring_chua}
\end{figure}

The network topology of the coupled Chua's circuits is shown in Fig.~\ref{fig:ring_chua}. This network is a mixture of the nonlocal circular ring topology, composed of $N$ nodes, and the star topology. The dynamics of the nodes in the ring is given by
\begin{align}
    \begin{split}
    \label{eq:diffusive}
    \dot{x}_{i} &=f_{x}(x_i,y_i,z_i) + k(x_{N+1}-x_{i}) + \frac{\sigma}{2rN}
    \sum_{j=i-rN}^{i+rN}(x_{j}-x_{i}),\\
    \dot{y}_{i} &=f_{y}(x_i,y_i,z_i)+\frac{\sigma}{2rN}\sum_{j=i-rN}^{i+rN}(y_{j}-y_{i}),
    \\
    \dot{z}_{i} &=f_{z}(x_i,y_i,z_i).   
    \end{split}
\end{align}
for $i=1,2,\ldots, N$ and the indices are considered modulo $N$. The dynamics of the central node is given by
\begin{align}
    \begin{split}
        \label{eq:central}
        \dot{x}_{N+1} &=f_{x}(x_{N+1},y_{N+1},z_{N+1})+\frac{k}{N}\sum_{j=1}^{N}(x_{j}-x_{N+1}),
        \\
        \dot{y}_{N+1} &=f_{y}(x_{N+1},y_{N+1},z_{N+1}),\\
        \dot{z}_{N+1} &= f_{z}(x_{N+1},y_{N+1},z_{N+1}). 
    \end{split}
\end{align}

In Eqs. \eqref{eq:diffusive} and \eqref{eq:central}, $x_i$, $y_i$, and $z_i$ are the dynamical variables of the $i^{\rm th}$ node, $k \geq 0$ is the star topology coupling intensity, and $\sigma \in [0, 1]$ and $r \in [0, 0.5]$ are the coupling intensity and the coupling radius of the ring topology, respectively. The functions $f_x$, $f_y$, and $f_z$ are the governing equations of the uncoupled node, which have the following form for the Chua's circuit:
\begin{align}
    \label{eq:chua}
    f_x &= \alpha\left\{y-x-\left[Bx+\frac{1}{2}(A-B)(|x+1|-|x-1|)\right]\right\},\\
    f_y &= x-y+z,\\
    f_z &= -\beta y,
\end{align}
with the parameters $A=-1.143$, $B=-0.714$, $\alpha=9.4$, and $\beta=14.28$. For these parameter values, the Chua's system shows its well-known double-scroll regime \cite{Matsumoto1984}. The size of the ring network is set to $N=150$, the coupling range to $r=1/3$, the ring coupling strength to $\sigma=0.68$, and the star coupling intensity to $k = 0.005$.

\begin{figure*}[t!]
    \centering
    \includegraphics[width=0.9\linewidth]{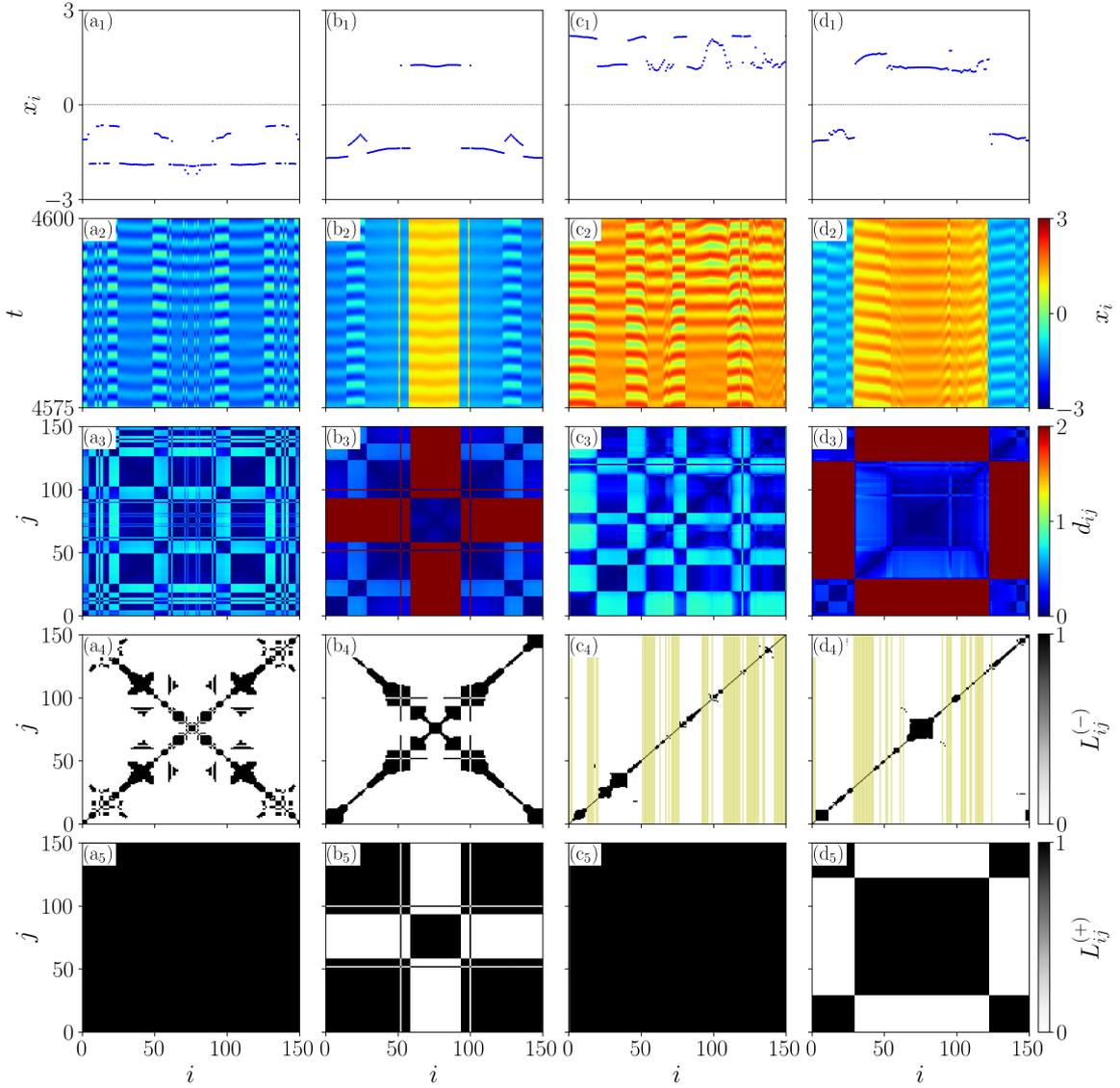}
    \caption{The snapshots of $x_i(t)$ at $t = 4600$ (1st row) and the spatio-temporal evolution of $x_i(t)$ (2nd row) of Eqs. \eqref{eq:diffusive} and \eqref{eq:central}, the distance matrix (3rd row), the binary distance matrix with a small threshold $\delta_s$ (4rd row), and the binary distance matrix with a large threshold $\delta_{\ell}$ (5th row) for (a) SW cluster state, (b) DW cluster state, (c) SW coherence-incoherence state, and (d) DW coherence-incoherence state for $N=150$, $\sigma=0.68$, $r=1/3$, $k=0.005$, $\delta_s=0.03$, $\delta_{\ell}=2.00$, $A=-1.143$, $B=-0.714$, $\alpha=9.4$, and $\beta=14.28$. The yellow vertical lines indicate the positions of the desynchronized nodes.}
    \label{fig:matrix}
\end{figure*}

\begin{figure*}[tb!]
    \centering
    \includegraphics[width=0.95\linewidth]{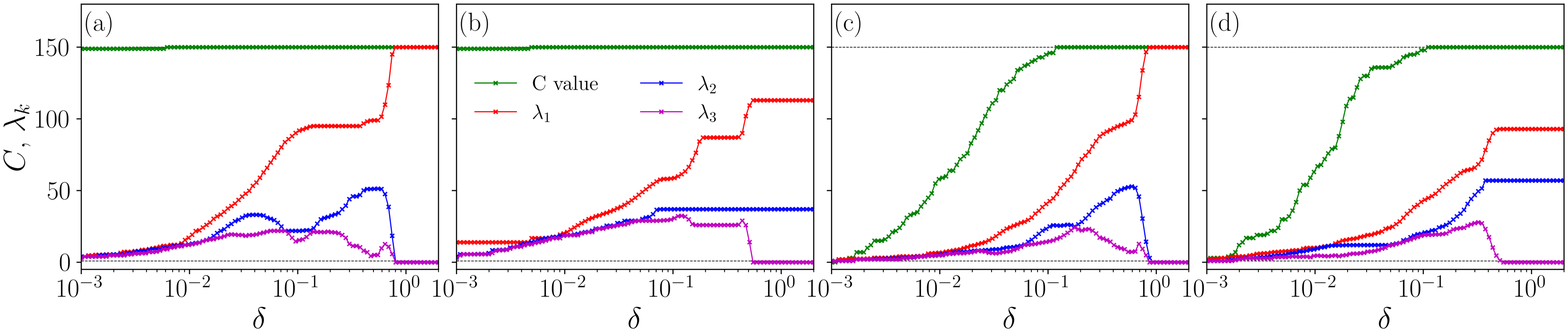}
    \caption{The $C$ value and the first three largest eigenvalues of the
    binary distance matrix as a function of the threshold $\delta$ of the Chua's network for (a)
    SW cluster state, (b) DW cluster state, (c) SW
    coherence-incoherence state, and (d) DW coherence-incoherence state for $N=150$, $\sigma=0.68$, $r=1/3$, $k=0.005$, $A=-1.143$, $B=-0.714$, $\alpha=9.4$, and $\beta=14.28$.}
    \label{fig:sum_eigen}
\end{figure*}

We consider parameters in which the oscillators stay in a single scroll attractor. These dynamic states have recently been called SW or DW \cite{Muni2020}. In the case of a SW dynamics, all the oscillators in the network belong to the same region of the phase space. In the DW, the dynamics of a fraction of oscillators is constrained to a region of the phase space, whereas the rest of them are constrained to a different, symmetric region. We remark that the uncoupled as well as coupled system \eqref{eq:diffusive}--\eqref{eq:central} of Chua's circuits possesses a central symmetry, \textit{i.e.}, the system is equivariant with respect to the transformation $x_i \mapsto -x_i$, $y_i \mapsto -y_i$, and $z_i \mapsto -z_i$ for all $i$. In particular, this guarantees a coexistence of two stable symmetric SW attractors in the uncoupled system. 
    
In order to observe these different dynamics, we consider all nodes to be at the unstable fixed point at the origin except for node $i = 1$, \textit{i.e.}, $(x_i, y_i, z_i) = (0, 0, 0)$ for $i = 2, 3, \ldots, N + 1$. The first row of Fig.~\ref{fig:matrix} shows the snapshots of $x_i$ at $t = 4600$ and the second row shows spatio-temporal evolution of $x_i(t)$ for four different initial conditions, \textit{i.e.}, four different values of $(x_1, y_1, z_1)$. In Fig.~\ref{fig:matrix}(a$_1$) we observe a SW cluster synchronisation (SW-CL) dynamics \cite{cluster_sync1,cluster_sync2,cluster_sync3}. In this case, the dynamics can be divided into two groups of oscillators with correlated dynamics, however the dynamics of all of them are roughly bound to the region $x_i<0$. In Fig.~\ref{fig:matrix}(b$_1$) we also observe the presence of groups of oscillators with correlated dynamics. In this second case, the dynamics of some of the oscillators  are constrained to a region given roughly by $x_i>0$, which is referred to as DW cluster synchronization (DW-CL). Fig.~\ref{fig:matrix}(c$_1$) displays a SW dynamics, in which some oscillators are not coherent with any of the others. This state is referred to as SW coherence-incoherence state (SW-CI). Figure \ref{fig:matrix}(d$_1$) exhibits a DW dynamics, where some of the oscillators are incoherent with the rest of the oscillators in the network. This last case is referred to as DW coherence-incoherence state (DW-CI). 

In order to quantify whether the state of the network is a coherence-incoherence or a cluster synchronization pattern we proceed in the following way:
\begin{enumerate}
    \item Sum the elements of each column of the binary matrix: $s_j = \sum_iL_{ij}^{s}$, for $i,j = 1, 2, \ldots, N$, \textit{i.e.}, excluding the central node.
    \item Apply the sign function on the sum of each column as $S_i = \sign(s_i - 1)$. This function assigns the value $1$ to the coherent nodes, as $s_i > 1$, and the value $0$ for the incoherent ones.
    \item Sum the elements of $S$: $C = \sum_{i = 1}^NS_i$. $C = N$ represents cluster synchronization and $C = 0$ shows that all the nodes are desynchronized. The coherence-incoherence state is characterized by intermediate values of $C$, $0 < C < N$, \textit{i.e.}, $C$ nodes belong to synchronized clusters and $N - C$ nodes are desynchronized.
\end{enumerate}

We apply the above methodology to the states shown in the first row of Fig.~\ref{fig:matrix}. The third row of Fig.~\ref{fig:matrix} shows the distance matrix $d_{ij}$ and the fourth row of Fig.~\ref{fig:matrix} displays the binary matrix $L_{ij}^{s}$ of the corresponding states for a small threshold $\delta_s = 0.03$. For the first two states (a) and (b), every node is coherent with at least one other node of the network and, hence, $C = N$ and we observe a cluster synchronization \cite{Parastesh2020}. For the remaining two states (c) and (d), it holds $C < N$, and we identify the incoherent nodes as the ones whose row/column of the binary matrix has all elements equal to zero but the diagonal one. We plot yellow stripes in the panels (c$_1$) and (d$_1$) to emphasize these nodes.


The performed procedure allows us to distinguish among cluster, coherence-incoherence, and desynchronized states. However, this method alone does not differentiate between SW and DW dynamics. In order to accomplish this, we first note from Fig.~\ref{fig:matrix} that for the DW dynamics some of the elements of the distance matrix are large when compared to the SW case. Thus, by changing the threshold $\delta$, we are able to differentiate between the SW and DW states. Indeed, the last row of Fig.~\ref{fig:matrix} shows the binary matrix $L_{ij}^{\ell}$ for a large threshold $\delta_{\ell} = 2.00$. For the SW cases, we see that all elements of the binary matrix equal $1$ whereas for the DW cases some of them are $0$. In this way, we can simply check if $L_{ij}^{\ell} = 0$ for some $i, j$, and in the affirmative case, the state is a DW state.
    
\begin{figure*}[tb]
    \centering
    \includegraphics[width=0.9\linewidth]{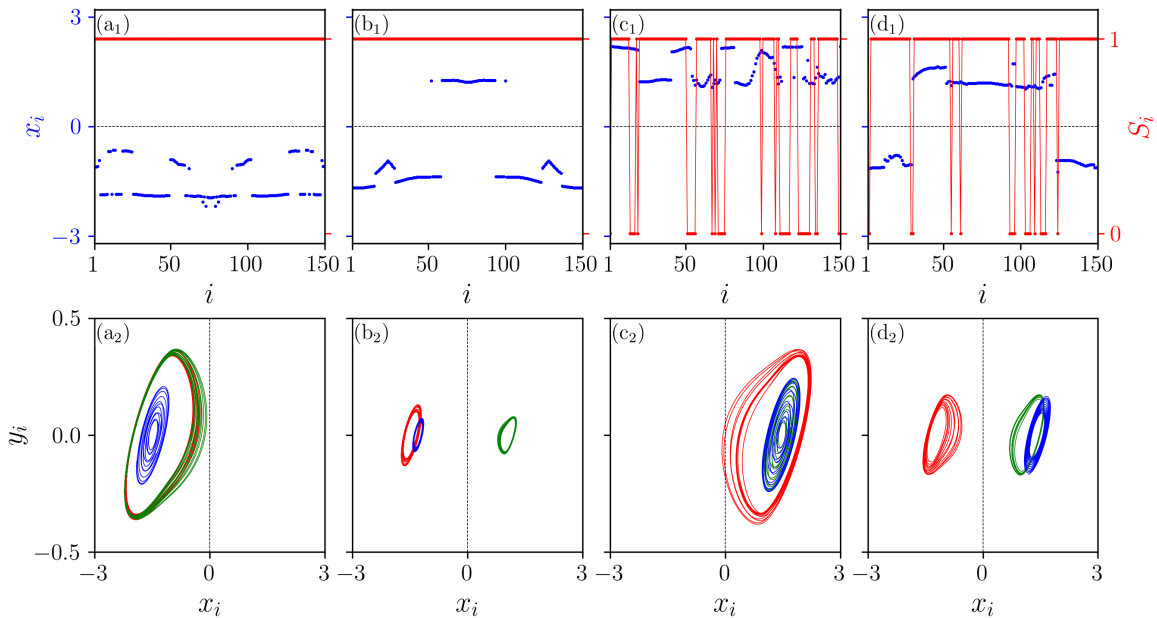}
    \caption{The snapshots of $x_i$ of the Chua's network at $t = 4600$, the vector $S_i$ (top), and the phase space for three different nodes (bottom) of a (a) SW cluster state, (b) DW cluster state, (c) SW coherence-incoherence state, and (d) DW coherence-incoherence state for $N=150$, $\sigma=0.68$, $r=1/3$, $k=0.005$, $\delta_s=0.03$, $A=-1.143$, $B=-0.714$, $\alpha=9.4$, and $\beta=14.28$.}
    \label{fig:snapshots}
\end{figure*}

To corroborate this methodology, we evaluate the $C$ value and the first three largest eigenvalues of the binary matrix (red, blue, and purple, respectively) as a function of the threshold $\delta$ (Fig.~\ref{fig:sum_eigen}). We verify that for small values of $\delta$, the $C$ value is equal to $N$ for the cluster states. However, it is less than $N$ for the coherence-incoherence states. Increasing $\delta$, the binary matrix has only one eigenvalue larger than $0$ in the SW cases, which equals the size of the matrix, and it has two in the DW cases. This fact confirms our previous statement that for SW dynamics $L_{ij}^{\ell} = 1$ for all $i,j$, as such a matrix has $N - 1$ zero eigenvalues and one equal to $N$. Therefore, in order to distinguish among cluster, coherence-incoherence, and desynchronized states, we use a small threshold $\delta_s=0.03$ and calculate $C$. Then, we check if the binary matrix for a large threshold $\delta_{\ell} = 2.00$ has all its elements equal to $1$. If so, the state is SW, and DW otherwise. Lastly, in Fig.~\ref{fig:snapshots}, we show in blue a snapshot of $x_i(t)$ of the nodes at $t=4600$. We separate the snapshots between SW and DW dynamics. In red, we plot the components of the vector $S_i$, which is used to identify coherent and incoherent regions in the network. And we also plot the projection of the dynamics of some sample nodes in the $(x,y)$ plane for each case.

\subsection{Cubic maps}

Another paradigmatic system that was reported to show SW and DW dynamics is a network of coupled cubic maps, described by the following equations \cite{shepelev2017chimera}:
\begin{equation}
    \label{eq:cubicmap}
    \begin{aligned}
        x_i(n + 1) &= f_i(n) + \frac{\sigma}{2rN}\sum_{j = i - rN}^{i + rN}\left[f_j(n) - f_i(n)\right],\\
        f_i(x) &= \left(\alpha x_i - x_i^3\right)\exp\left(-\frac{x_i^2}{\beta}\right),
    \end{aligned}
\end{equation}
where $i = 1, 2, \ldots, N$, $N$ is the size of the network, $\sigma \in [0, 1]$ and $r \in [0, 0.5]$ are the coupling strength and coupling radius, respectively, and $\alpha$ and $\beta$ are the control parameters of the map. The cubic map can be thought of as an analog of the Chua's circuit with chaotic dynamics for discrete-time systems and the cubic map is the simplest map to present bistable dynamics \cite{cubicmap1}. We remark here again the central symmetry $x_i \mapsto -x_i$ leading to the emergence of symmetric coexisting attractors. 

\begin{figure*}[tb]
    \centering
    \includegraphics[width=0.9\linewidth]{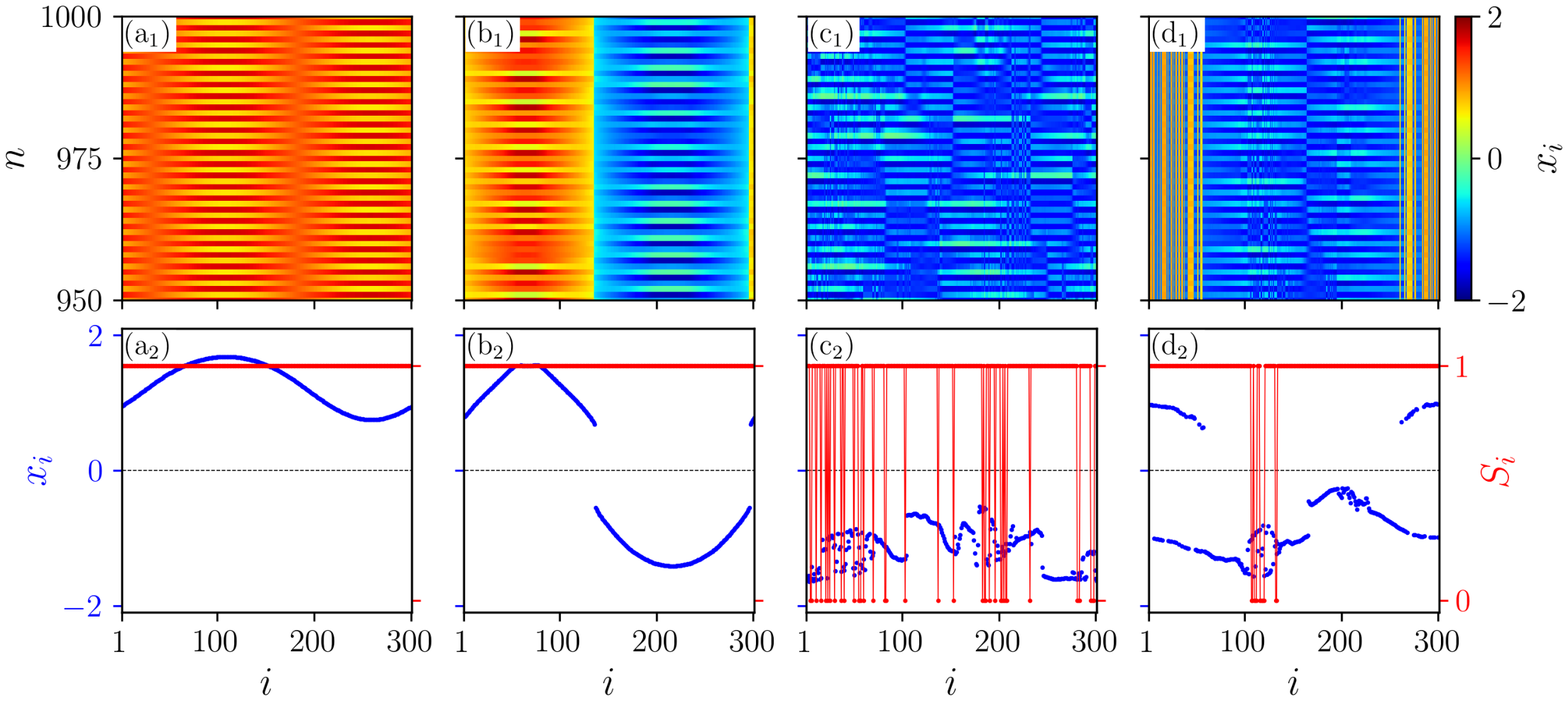}
    \caption{The spatio-temporal evolution of $x_i(n)$ of Eqs. \eqref{eq:cubicmap} (top), the snapshot of $x_i$ at $n = 1000$ in blue, and the components of the vector $S_i$ in red (bottom) for a (a) SW cluster with $(\sigma, r) = (0.8, 0.275)$, (b) DW cluster with $(\sigma, r) = (0.6, 0.2)$, (c) SW coherence-incoherence with $(\sigma, r) = (0.45, 0.2)$ and (d) DW coherence-incoherence with $(\sigma, r) = (0.45. 0.2)$. The others parameters are $\alpha = 3$, $\beta = 10$, $N = 300$, $\delta_s = 0.06$ and $\delta_{\ell} = 2.00$.}
    \label{fig:cubicmap}
\end{figure*}

For the control parameters we set $\alpha = 3$, $\beta = 10$, that correspond to a chaotic attractor of the map \cite{shepelev2017chimera}, and $N = 300$ for the network size. As the coupling strength and coupling radius change, different regimes are observed in the system \cite{shepelev2017chimera}, such as complete chaotic synchronization, partial coherence with SW and DW structures, and SW and DW coherence-incoherence. By applying our methodology using the parameters $(\sigma, r)$ reported in \cite{shepelev2017chimera}, we are indeed able to distinguish cluster synchronization from coherence-incoherence states and SW from DW as well. In Fig.~\ref{fig:cubicmap} are shown the spatiotemporal evolution of $x_i$ of the system \eqref{eq:cubicmap} and in blue a snapshot of $x_i$ at $n = 1000$, and in red we plot the components of the vector $S_i$. The parameters are (a) $(\sigma, r) = (0.8, 0.275)$, (b) $(\sigma, r) = (0.6, 0.2)$, (c) and (d) $(\sigma, r) = (0.45, 0.2)$, and we choose random initial conditions for all the nodes in the interval $x \in [-2, 2]$. For the coherence-incoherence detection we use $\delta_s = 0.06$ and for the SW/DW distinction $\delta_{\ell} = 2.00$.

For the first two states, Figs. \ref{fig:cubicmap}(a) and \ref{fig:cubicmap}(b), as all elements of $S_i$ equal to $1$, we get $C = N$, that corresponds to cluster synchronization. Whereas for the last two states, we observe $C < N$, indicating partial incoherence in the network. By increasing the threshold, as mentioned before, we are also able to distinguish between the SW states from the DW states.

Therefore, this methodology is a powerful tool in the detection of different dynamical behavior in networks of coupled oscillators. We would like to stress that the methodology proposed by Parastesh \textit{et al.} \cite{Parastesh2020} using the local order parameter matrix also gives similar results (not shown). However we use the distance matrix as it is simpler and it has a more direct interpretation in terms of synchronized/desynchronized states.

In the next section we employ our methodology to study how changes in the star coupling strength affects the spatial states of the network of $N + 1$ Chua circuits, given by Eqs. \eqref{eq:diffusive} and \eqref{eq:central}.


\section{Coexistence of different states and parameter dependence}
\label{sec:result}

Here we study how spatio-temporal patterns change with varying parameters. Also, perturbed initial conditions can lead to completely different dynamics, due to complex structures of the basin of attraction \cite{Va18,Va20}. As a first analysis of the influence of the central node on the ring network dynamics, we calculate the basin stability (BS) \cite{basinstability1,basinstability2} for each of the four types of states in our network and its change when $k$ is varied.

To estimate the value of BS, we consider $1000$ random initial conditions in the interval $[-1,1]$ for $x_1(0)$ and $y_1(0)$, and calculate the fraction of these points that converge to each of the four observable dynamics in the network. In that way, BS provides the proportion of the volumes of the basins of attraction for a given interval \cite{Menck13,Menck14,Schultz14}. 
    
\begin{figure}[tb]
    \centering
    \includegraphics[width=0.95\linewidth]{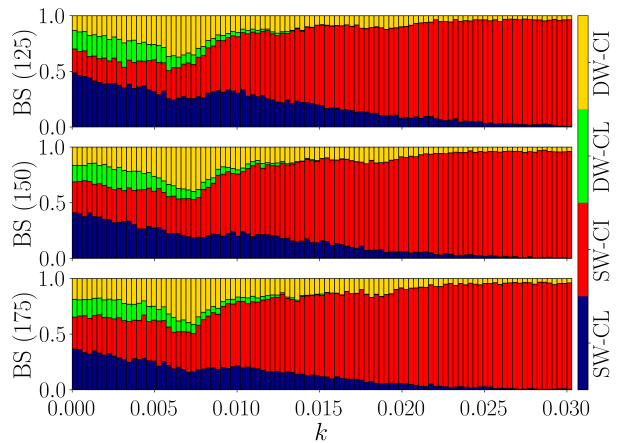}
    \caption{Fraction of initial conditions that converges to each state as a
    function of $k$ for different network sizes with $\sigma=0.68$, $r=1/3$,
    $\delta_s=0.03$, $\delta_{\ell}=2.00$, $A=-1.143$, $B=-0.714$, $\alpha=9.4$,
    and $\beta=14.28$.}
    \label{fig:bs}
\end{figure}

In Fig.~\ref{fig:bs}, we plot the value of BS obtained for $k$ in the interval $[0.0,0.03]$ for three distinct network sizes $N=125,150$, and $175$. The color code used is as follows: blue for SW cluster states, green for DW cluster states, red for SW coherence-incoherence states, and yellow for DW coherence-incoherence states. From the graphs, we see that for $k=0$ the basins of all the states have approximately the same volume. By increasing the value of $k$, we verify that the basin of DW-CL begins to diminish so that from $k=0.015$, those states are not observed in the network. We may also note that this happens independently of the network size $N$. By increasing $k$ even further the basins of both SW-CL and DW-CI also get smaller, in a way that there is a dominance of SW-CI. The influence of the central node in the ring network dynamics is such that it privileges the SW-CI and suppresses the other types of dynamics observed for $k=0$.

\begin{figure}[tb]
    \centering
    \includegraphics[width=0.95\linewidth]{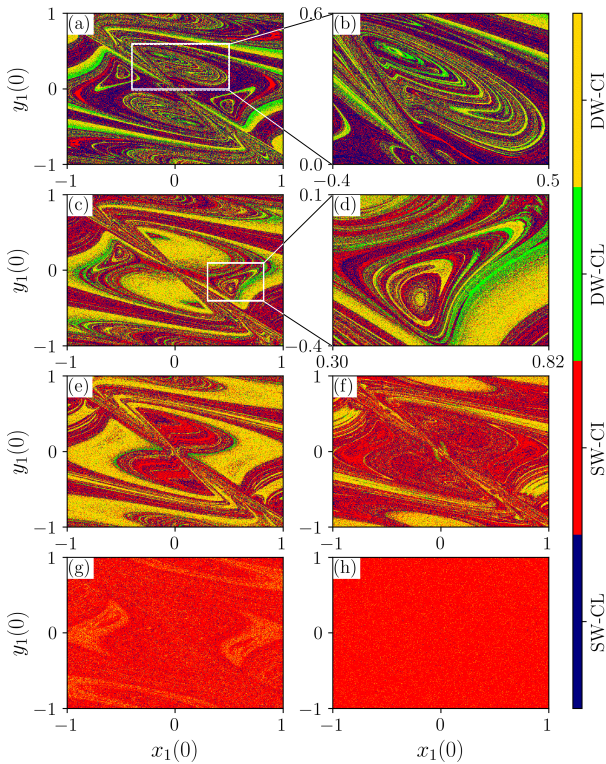}
    \caption{Basins of attraction of the ring-star network of Chua's circuits with
    $N=150$, $\sigma=0.68$, $r=1/3$, $\delta_s=0.03$, $\delta_{\ell}=2.00$,
    $A=-1.143$, $B=-0.714$, $\alpha=9.4$, $\beta=14.28$, and (a) and (b) $k=0.0$, (c) and (d)
    $k=0.005$, (e) $k=0.0075$, (f) $k=0.01$, (g) $k=0.02$, and (h) $k=0.03$.}
    \label{fig:basins}
\end{figure}

In Fig.~\ref{fig:basins}, we plot the basins of attraction for different values of $k$. To obtain the basins, we initially consider all nodes at the unstable fixed point at the origin, $(x_j, y_j, z_j)=(0,0,0)$ for $j=2,\dots,N+1$, and vary $x$ and $y$ of the node $i=1$. The basins are constructed using a grid of $540\times 540$ initial conditions. The color of each point is given according to the dynamical state of the network after a total integration time $t=4600$. Without the presence of the central node (Figs.~\ref{fig:basins}(a) and \ref{fig:basins}(b)), the basins show a very complex structure, with the basins of each of the states occupying a significant part of the given region. For $k=0.005$ (Figs.~\ref{fig:basins}(c) and \ref{fig:basins}(d)), we perceive a very significant change in the structure of the basins, even though the change in BS is more subtle. Moreover, the magnifications of the regions indicated by the white rectangles (Figs.~\ref{fig:basins}(b) and \ref{fig:basins}(d)) show that the boundaries of the basins resemble fractal and riddled structures \cite{rid1, rid2}. In Fig.~\ref{fig:basins}(e), we consider $k=0.0075$, for this value the BS of DW-CI is larger (Fig.~\ref{fig:bs}), and we find a significant decrease in the size of the basin of DW-CL. In Figs. \ref{fig:basins}(f), \ref{fig:basins}(g), and \ref{fig:basins}(h), we increase the central node coupling to $k=0.01$, $k=0.02$, and $k=0.03$, respectively. We see from these panels that the blue, green and yellow regions almost vanish, and mostly the basin of SW-CI remains, although riddled with points from the other basins.


\section{Conclusions}
\label{sec:concl}

We have analyzed the influence of a central node in a network of nonlocally coupled systems. We first characterized the dynamical states in a network of Chua's oscillators and in a network of cubic maps. By considering two values of the threshold to define the binary matrix $\mathbf{L}$, we were able to separate the dynamics among SW cluster states, SW coherence-incoherence states, DW cluster states, and DW coherence-incoherence states. We have also tested whether the maximum distance, instead of the mean distance, in the definition \eqref{eq:distmat}, should be a better choice. However our simulations indicated that is not the case, as the maximum distance detected some false positive DW states (not shown).
    
The estimation of basin stability allowed us to verify how the probability of obtaining each state changes by increasing the strength of the central node coupling $k$. This analysis showed that the influence of the central node causes the suppression of some dynamical states, increasing the size of SW-CI basin. Although it allows us to see the fraction of initial conditions that converge to each state, BS alone can not tell us much about the structures in the basins. From Fig.~\ref{fig:basins}, we see that the structures are very complex and the magnifications (Figs.~\ref{fig:basins}(b) and \ref{fig:basins}(d)) show that the basins have riddled structures. From these figures, we verify that the changes in the proportions of each state happens simultaneously with drastic changes in their boundaries. For larger values of $k$, the basin of SW-CI dominates and is riddled with points of the other states.

An interesting point for future studies could be a more detailed analysis of a coexistence of different coherence-incoherence patterns. Indeed, our method allows to distinguish four big classes of patterns: SW and DW clusters, SW and DW coherence-incoherence patterns. However, it is obvious that there can be a high multistability within each class of these solutions, which makes the problem even more challenging. The situation can become even more complicated if the dynamics of the node has a more complicated symmetry than a central symmetry considered in our work. For example, $S^n$ symmetry may lead to a coexistence of $n$ symmetric attractors in the node's dynamics which could lead to $n$-well states. 
    
\section*{Data availability} 

The data that support the findings of this study are available from the corresponding author upon reasonable request.
    
\section*{Declaration of competing interest}
    
The authors declare that they have no known competing financial interest or personal relationships that could have appeared to influence the work reported in this paper.

\section*{Acknowledgements}

We wish to acknowledge the support of the Arauc\'aria Foundation, the Coordination of Superior Level Staff Improvement (CAPES) and the National Council for Scientiﬁc and Technological Development (CNPq). S.S.M acknowledges the School of Fundamental Sciences doctoral bursary funding, Massey University during this research. S.S.M acknowledges the use of NeSI (New Zealand e-Science Infrastructure) HPC computing facilities during this research. V. S acknowledges Fatemeh Parastesh for the help in understanding the chimera states detection using eigenvalue decomposition. S.Y. was supported by the German Research Foundation DFG, Project No. 411803875. J.K. has been supported by the Alexander von Humboldt Polish Honorary Research Scholarship 2020 of the Fundation for Polish Science. We would also like to thank the 105 Group Science (\url{www.105groupscience.com}) for fruitful discussions.
    
\bibliographystyle{ieeetr}
\bibliography{references}
    
\end{document}